\documentclass[9pt,twocolumn,twoside]{opticajnl}
\journal{opticajournal} 

\setboolean{shortarticle}{true}

\usepackage{lineno}
\usepackage[toc,page]{appendix}

\title{Few photons probe third-order nonlinear properties of nanomaterials in a plasmonic nanocavity}

\author[1]{Anupa Kumari}
\author[1]{Mohammadreza Aghdaee}
\author[1,2]{Mathis Van de Voorde}
\author[1*]{Oluwafemi S. Ojamabati}

\affil[1]{Faculty of Science and Technology,  MESA+ Institute, University of Twente, Enschede, 7522NB, the Netherlands}
\affil[2]{Institute of Microstructure Technology, Karlsruhe Institute of Technology, Karlsruhe, Germany}
\affil[*]{Corresponding author: o.s.ojambati@utwente.nl}

\begin{abstract}
Quantification of nonlinear optical properties is required for nano-optical devices, but they are challenging to measure on a nanomaterial. Here, we harness enhanced optical fields inside a plasmonic nanocavity to mediate efficient nonlinear interactions with the nanomaterials. We performed reflection Z-scan technique at intensity levels of kWcm$^{-2}$, reaching down to two photons per pulse, in contrast to GWcm$^{-2}$ in conventional methods. The few photons are sufficient to extract the nonlinear refractive index and nonlinear absorption coefficient of different nanomaterials, including perovskite and Au nano-objects and a molecular monolayer. 
This work is of great interest for investigating nonlinear optical interactions on the nanoscale and characterizing nanomaterials, including fragile biomolecules.

\end{abstract}

\setboolean{displaycopyright}{false} 

\begin{document}

\maketitle


Nonlinear optical materials are essential in advanced photonics, serving as building blocks for devices such as optical limiters, saturable absorbers, and optical switches \cite{dini2016nonlinear, wang2021highly, wang2019saturable, min2008all}. Moreover, on the nanoscale, there is also a growing demand for the exploitation of nonlinear optical devices in photonic circuits and chips \cite{sirleto2023introduction,koenderink2015nanophotonics}. An accurate determination of the nonlinear optical properties is, therefore, valueable. Particularly, these applications require quantification of nonlinear susceptibilities $\chi^{(N)}$, $N=3,5,..$, with real and imaginary parts related to the nonlinear refractive index and nonlinear absorption coefficient, respectively.

However, there is still a knowledge gap in determining the nonlinear optical properties of nanomaterials. This knowledge gap is due to the absence of a straightforward experimental method that efficiently focuses optical fields on the nanomaterial. There are various demonstrations of nonlinear optical effects on nanomaterials including four-wave mixing \cite{xiang2016nanoparticle,Wang_AdvOptPhoton_2011}, high harmonic generation \cite{singhal2010study,bonacina2020harmonic}, two-photon absorption \cite{shen2016two, lu2022two}, optical Kerr effect \cite{tomita2006magneto, huang2023weak}, nonlinear spectroscopy \cite{obermeier2018nonlinear, schumacher2011nanoantenna}, and supercontinuum generation \cite{driben2009low,besner2006fragmentation}. Despite the popularity of these observations, nonlinear optical properties are inferred indirectly, and the data interpretation is complicated \cite{samoc1998femtosecond}. 

Z-scan method is a popular experimental technique to directly determine the third-order nonlinear refractive index  ($n_2$) and nonlinear absorption coefficient ($\beta$) of thin films \cite{sheik1990sensitive, van1997z}. A sample is scanned through an optical focus, and changes in transmitted or reflected intensities are compared with a theoretical model. The fitting parameters of the model are $n_2$ and $\beta$. 
The Z-scan technique has two main limitations: First, it is not yet suitable for characterizing single nanoscale materials because the incident field is focused on area of about tens of $\mu \mathrm{m} ^2$. Thus, the intensity is inefficiently distributed on a single nanoparticle. Second, the existing Z-scan measurements still require high intensities of the order of GWcm$^{-2}$ \cite{tian2024dispersion}. Such high intensities pose a significant risk of damaging nanomaterials, particularly delicate structures like biomaterials. 

Plasmonic nanostructures are a possible solution to overcome the challenges of measuring nonlinear optical properties. The nanostructures localize optical fields in a sub-diffraction-limited spot, due to induced collective oscillations of free electrons \cite{maier2007plasmonics}. The localized fields can efficiently concentrate the intensity on a single nanomaterial as well as enhance nonlinear optical interactions due to substantial field enhancements \cite{kauranen2012nonlinear, butet2015optical,Bouhelier_PhysRevLett_2003,Lippitz_NanoLett_2005,li_nonlinear_2017,ojambati2020efficient}. However, there are still no measurements of the nonlinear properties of nanomaterials using plasmonic nanostructures, especially in combination with Z-scan.

In this Letter, we measure the third-order nonlinear optical properties, namely $n_2$ and $\beta$, of nanoscale materials inside a plasmonic nanocavity. The plasmonic nanocavity is made of Au nanoparticle that is separated by the material to be characterized from a Au film. Optical fields are enhanced inside the nanocavity \cite{wang_NatRevChem_2022, baumberg_extreme_2019}, thus providing a suitable environment for efficient nonlinear optical light-matter interactions. We demonstrate that the highly efficient nanocavity enables low-intensity illumination down to a few photon levels.



\begin{figure}[htbp]
\centering
\includegraphics[width=\linewidth]{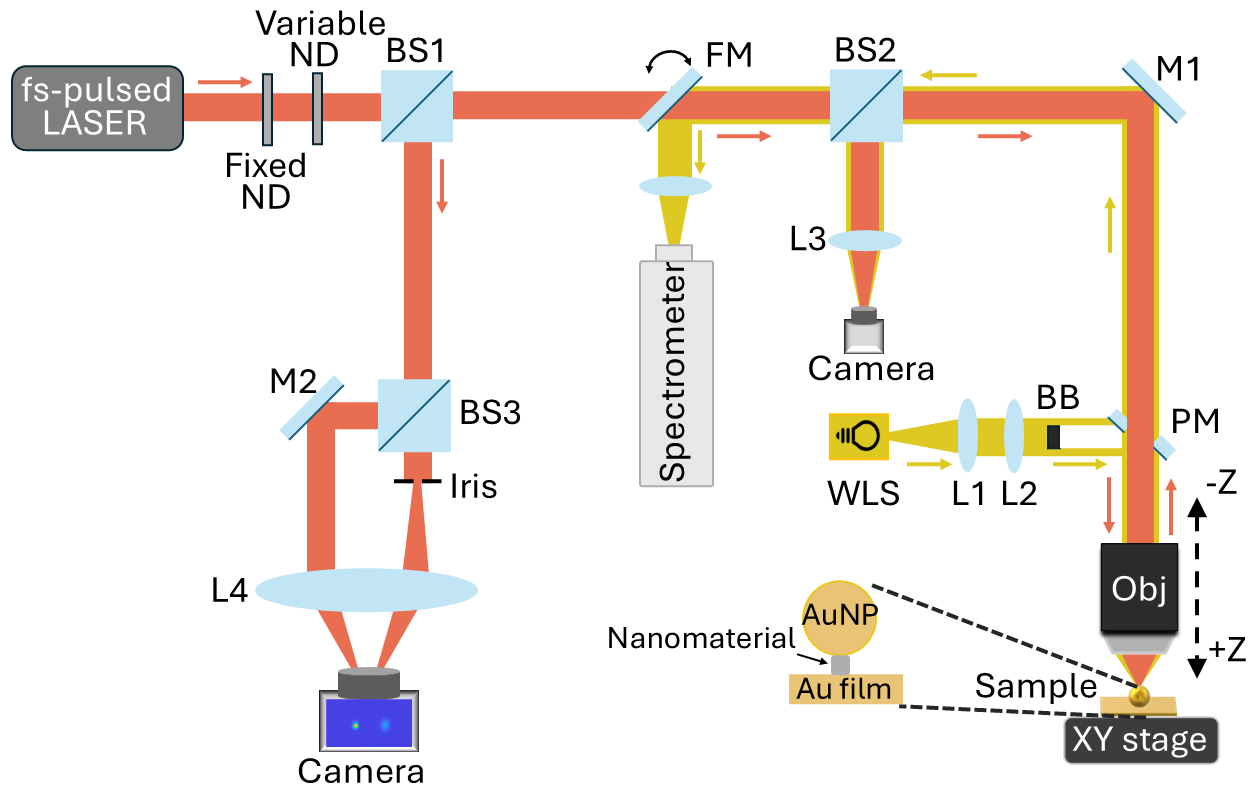}
\caption{Schematic of reflection Z-scan experimental setup. L1, L2, L3, L4: lens; M1, M2: mirror; PM: perforated mirror; ND: neutral density filter; BS1, BS2, BS3: beam splitter; FM : flipping mirror; WLS: white light source; BB: beam blocker. }
\label{fig1}
\end{figure}
 
To determine the third-order nonlinear properties, we used the reflection Z-scan (RZ-scan) technique. The principle of our measurement is as follows: We measure the reflected light in two configurations, open aperture (OA) and closed aperture (CA), as the sample moves through the profile of a focused laser beam. In the OA configuration, the reflected light is collected by a lens and measured by a detector. The OA configuration provides information about $n_2$ due to the nonlinear photo-induced modification of the reflection coefficient of the surface. Whereas for CA, only a part of the beam passing through an iris is collected by the detector. The CA captures the nonlinear phase change of the reflected beam due to the nonlinear absorption. No nonlinear effects appear when the sample is positioned far from the focal plane, and therefore the reflected laser beam in this region is constant.

In the experiment, we employ a combined RZ-scan setup and a darkfield microscropy-spectroscopy setup (Figure \ref{fig1}) (for more details, see the Supplementary material). A femtosecond pulsed laser delivers 120 fs optical pulses with a tunable wavelength in the near-infrared region. The pulses have a repetition rate of 76 MHz and an average output power of 1 W. A combination of variable and fixed neutral density filters control the intensity that illuminates the sample. The laser pulse is focused onto the sample using a brightfield/dark-field microscope objective with a numerical aperture of 0.9. The microscope objective translates in the z-direction, bringing the sample through the optical focus. The sample is mounted on a XY translation stage to be able to measure multiple nanocavities. The darkfield microscopy-spectroscopy setup images the nanoparticles on the Au film and measures their extinction spectra, to reveal the resonance of the individual plasmonic cavity. 


We investigate three nanomaterials inside the plasmonic nanocavity: (1) 10 nm Au nano-object, (2) 6.5 nm Yb$^{3+}$:CsPbCl$_3$ perovskite nano-object, and (3) monolayer of methylene blue molecule, that is 0.9-nm thick. (For details about the sample fabrication and characterization, see Supplementary Information). These samples enable a comparison of the nonlinear effects inside the nanocavity at different gap sizes and nonlinear optical properties. A bare Au film without a nanomaterial serves as a reference sample. For all the samples, we use an atomically flat Au film with a thickness of 40 nm instead of a thicker film (> 100 nm). The thin Au film allows a considerable scattered intensity from the nanocavity and reduces reflection from the Au film. Gold nanoparticles (80 nm diameter) were dropcasted on the materials on the Au film to form the nanocavity.


\begin{figure}[htbp]
\centering
\includegraphics[width=0.85\linewidth]{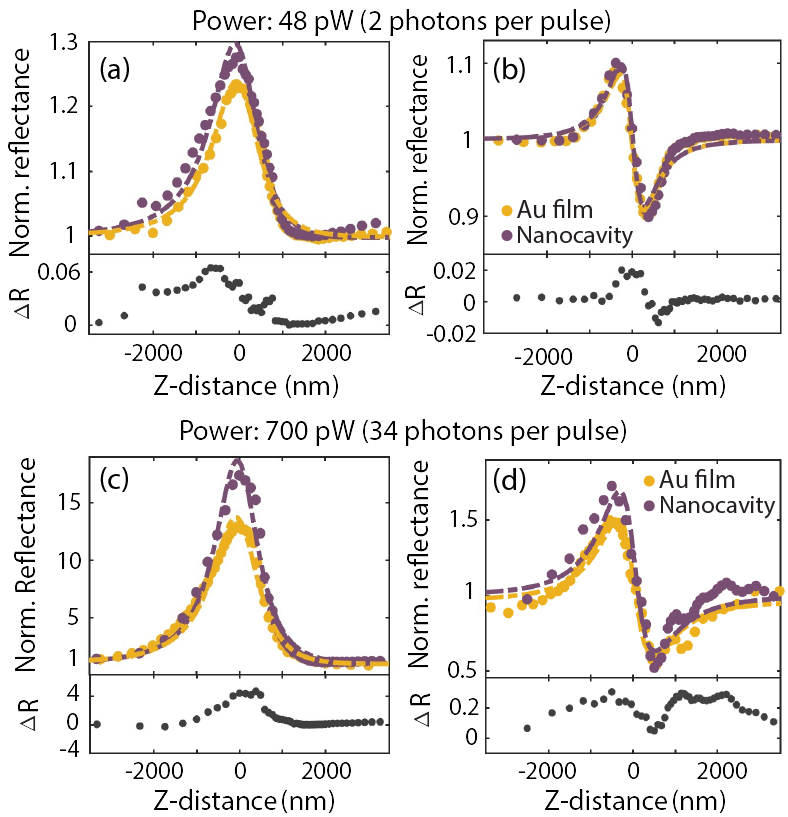}
\caption{ Normalized reflectances $\textit{vs}$ z-distance of the objective for open (a,c) and closed aperture (b,d) measurements on the nanocavity (purple) and on the gold film (yellow) at 48 pW and 700 pW. The dotted lines represent the theoretical fit. Insets: Differential reflectance obtained by subtracting the Au film counts from the nanocavity counts for each graph, is indicated with a black marker.}
\label{fig2}

\end{figure}

We probe the Au nano-objects in the nanocavity at two lasers powers of 48 pw and 700 pW, corresponding to 2 and 34 photons per pulse on average, respectively, at 750 nm. The OA Z-scan trace shows that the reflectance has a peak around the focus (Figure \ref{fig2}a). On the other hand, the CA Z-scan trace shows a peak-to-valley signature, meaning that the reflectance increases in the prefocal region and decreases in the postfocal region. Conventionally, the Z-scan traces are normalized to the reflected intensity at large z-distances away from the focus, which is the linear regime \cite{van1997z, liu2001theoretical}. When the sample approaches the focal plane, the laser intensity increases and the nonlinear effects start occurring. 
Both configurations agree with expected theoretical results, as we explain below. For the OA, the reflectance peak is attributed to positive values of the nonlinear refractive index ($n_{2} > 0$) due to changes in the reflection coefficient \cite{ganeev2005reflection}. The CA Z-scan trace follows a peak-to-valley signature, indicating a positive change in the nonlinear phase that results in positive values for the absorption coefficient ($\beta > 0$).

An interesting point to note here is the possibility of performing these measurements at extremely low photon count levels, down to an average of 2 photons per pulse.  This result highlights the sensitivity of our experimental set-up to the optical excitation, which is about $10^6$ more efficient than typical Z-scan techniques that uses intensity of the order of GWcm$^{-2}$. In comparison to the reflection from a bare Au film, there is an observable contribution from the nanocavity. We quantify this contribution by the difference in the reflectance ($\Delta$R) between gold film and nanocavity (insets in Figures \ref{fig2}). Both the OA and CA measurements show a non-zero $\Delta$R that increases by almost 10 times between the two powers. 

\begin{figure}[h!]
\centering
\includegraphics[width = \linewidth]{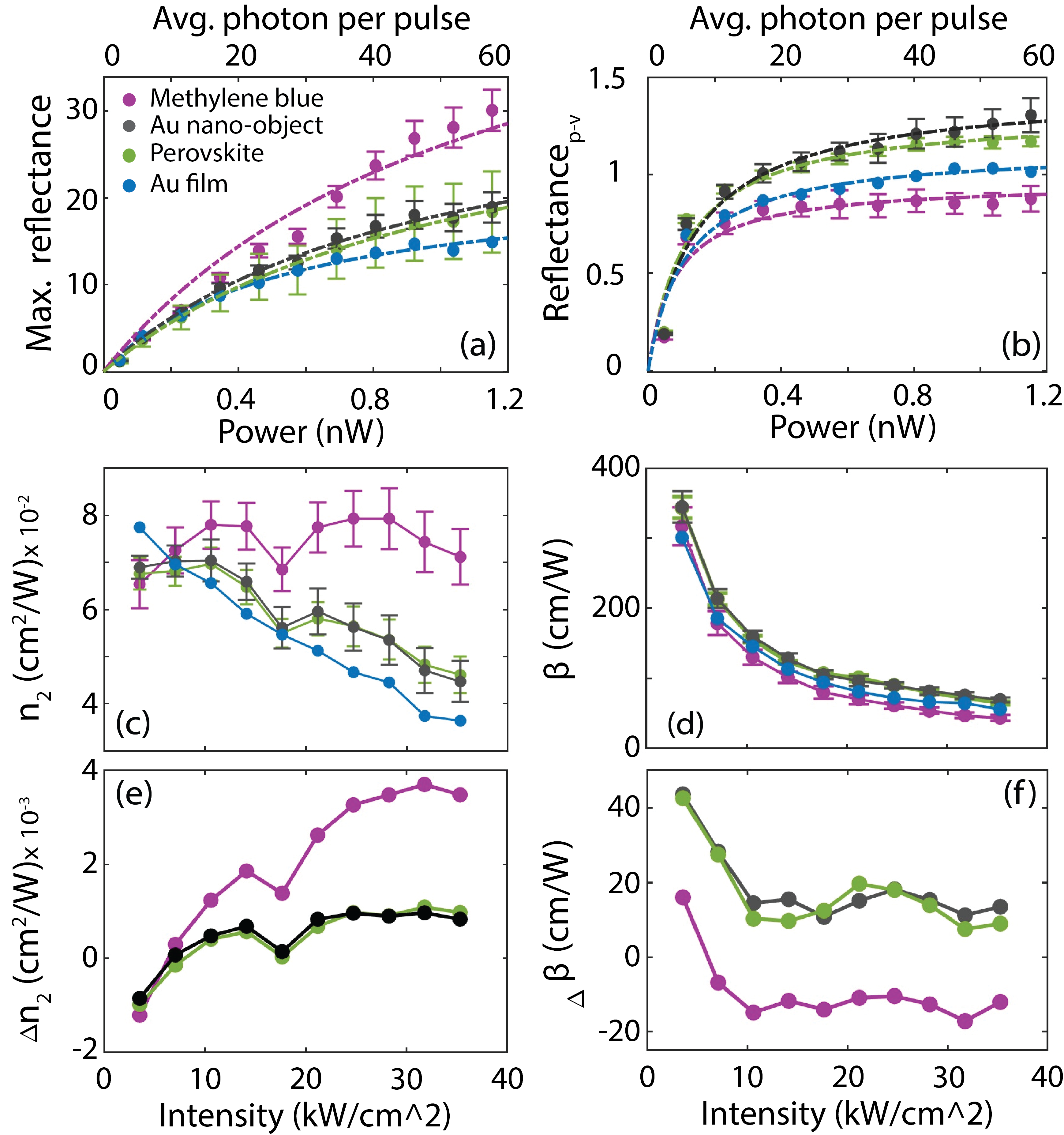}
\caption{(a,b) Maximum reflectance and the peak-to-valley (P2V) reflectance $\textit{vs}$ average pulse power, extracted from the open aperture and closed aperture measurements. The dotted line represents the intensity saturation fitting curve(c,d) Extracted nonlinear refractive index $n_2$ and the nonlinear absorption coefficient $\beta$ $\textit{vs}$ intensity. (e,f) Nonlinear refractive index difference $\Delta n_2$ and nonlinear absorption coefficient difference $\Delta \beta$ between the nanocavity and the Au film. The error bar is the standard deviation of the distribution for an average of 25 nanocavities per sample.}
\label{fig3}
\end{figure}

We performed power-dependent RZ-scan measurement on 25 nanocavities, on average, for the three different sample. We extract the maximum reflectance and the peak-to-valley reflectance from the Z-scan traces.  Generally, for both OA and CA, the maximum reflectance and the peak-to-valley reflectance increase with the average pulse power (Figures \ref{fig3} a,b). 
In addition, both measurements show saturation behaviors, albeit at different saturation powers of 1 nW and 0.1 nW on average for OA and CA, respectively. For the OA measurements, the methylene blue layer has the highest reflectance than the Au and perovskite nano-objects, and implying that there is a higher $n_2$. On the other hand, the peak-to-valley reflectance of the methylene blue layer is the lowest in CA. As CA is attributed to the nonlinear absorption happening in the material, one can say that the Au and perovskite nano-objects result in a higher nonlinear absorption than when the methylene blue layer is inside the gap. In CA, the reflectance values are generally lower than OA measurements because only a part of the beam ($20\%$) passes through the iris.

For the quantitative estimation of the $n_{2}$, we fit the OA Z-scan traces with a theoretical model \cite{petrov1996reflection, ganeev2008single} that describes the normalized reflectance $R$ (see SI)

 \begin{equation}
R(x) = 1+ \frac{r\,I_0\, n_{2}}{(x^2+1)(x^2+9)} + \mathcal{O}(k^{\prime}_2) \, ,
\label{eq1}
\end{equation}
$r$ is the relative change in the reflection coefficient due to the nonlinear optical effects, $I_0$ is the laser intensity on the sample and $x = z/z_{0}$, $z_0$ is the Rayleigh length. A higher-order correction term is included as $\mathcal{O}(k^{\prime}_2)$, where $k^{\prime}_2$ is a small factor that accounts for nonlinear absorption. 
Similarly, we model the measured CA Z-scan traces with the equation 

\begin{equation}
R(x) = 1- \frac{4\,r\,I_0\,x\,k_{2}}{(x^2+1)(x^2+9)} + \frac{r^2\,{I_0}^2\,k_2^2}{(x^2+1)(x^2+9)} + \mathcal{O}( n^{\prime}_2) \, ,
\label{eq2}
\end{equation}
that also includes higher-order correction terms that include a small contribution from nonlinear refraction, accounted for by $n^{\prime}_2$.
We extract $n_2$ and $k_2$ by fitting the Eqs (2) and (3) respectively to the Z-scan traces and calculate $\beta = 2\pi k_2/\lambda$, where $\lambda$ is the wavelength, for all input intensities and the materials. 
The extracted $\beta$ for the Au film is comparable to the result ($2\times10^{-3} \,$cm/W at 110 MWcm$^{-2}$) in the literature \cite{Smith_JAppPhys_1999}, after accounting for the differences in the intensity and sample thickness. This agreement shows the efficacy of our extraction procedure. The $n_2$ for Au film displays a downward linear trend with increasing intensity. The nanocavity samples show a slight oscillatory behaviour with a noticeable dip around $17 kW/cm^2$. Similar to the Au film, the nano-objects also exhibit a downward trend, suggesting an influence from the Au film. 

To obtain the $n_2$ values specific to the materials inside the nanocavity, we calculate the differential nonlinear refractive index by subtracting the $n_2$ values of the Au film from those of the materials. After this subtraction, all the materials show an increasing trend followed by saturation (Figure \ref{fig3}e). The methylene blue layer demonstrates up to about four times higher $\Delta n_2$ than the nano-objects. The values of the $\Delta n_2$ of the Au and perovskite nano-objects match quite well. We attribute the $n_2$ values to the different optical fields confined on the nanomaterials by the nanocavity. Simulation result show that the maximum field enhancement inside the gap is 330 and 11 compared at 750 nm for the methylene blue layer and the nano-objects, which have larger gap sizes (Fig. S6). However, the extracted $n_2$ is also determined by the position of the nano-objects in the gap, where the field enhancement transversely varies. These findings qualitatively indicate that optical fields in the gap result in different $n_2$ values due to modification of the reflection coefficients in the gap.


In the closed aperture (CA) measurements, $\beta$ decreases with intensity for all the materials (Figure \ref{fig3}d). We also calculated the differential nonlinear refractive index ($\Delta\beta$) by subtracting the $\beta$ values of the Au film from those of the materials. The data reveal that the methylene blue layer has negative $\Delta\beta$ values (Figure \ref{fig3}f), indicating saturable absorption or population inversion, which is consistent with a previous observation of lasing of methylene blue in the nanocavity \cite{ojambati2024few}. In contrast, $\beta$ is positive for both the nano-objects, and their values overlap at higher intensities. The positive $\beta$ implies that there is nonlinear absorption (e.g. two-photon absorption), which can occur due to broad electronic states at shorter wavelengths (Fig. S3 in SI) for both the perovskite and Au nano-object.


\begin{figure}[htbp]
\centering
\includegraphics[width =\linewidth]{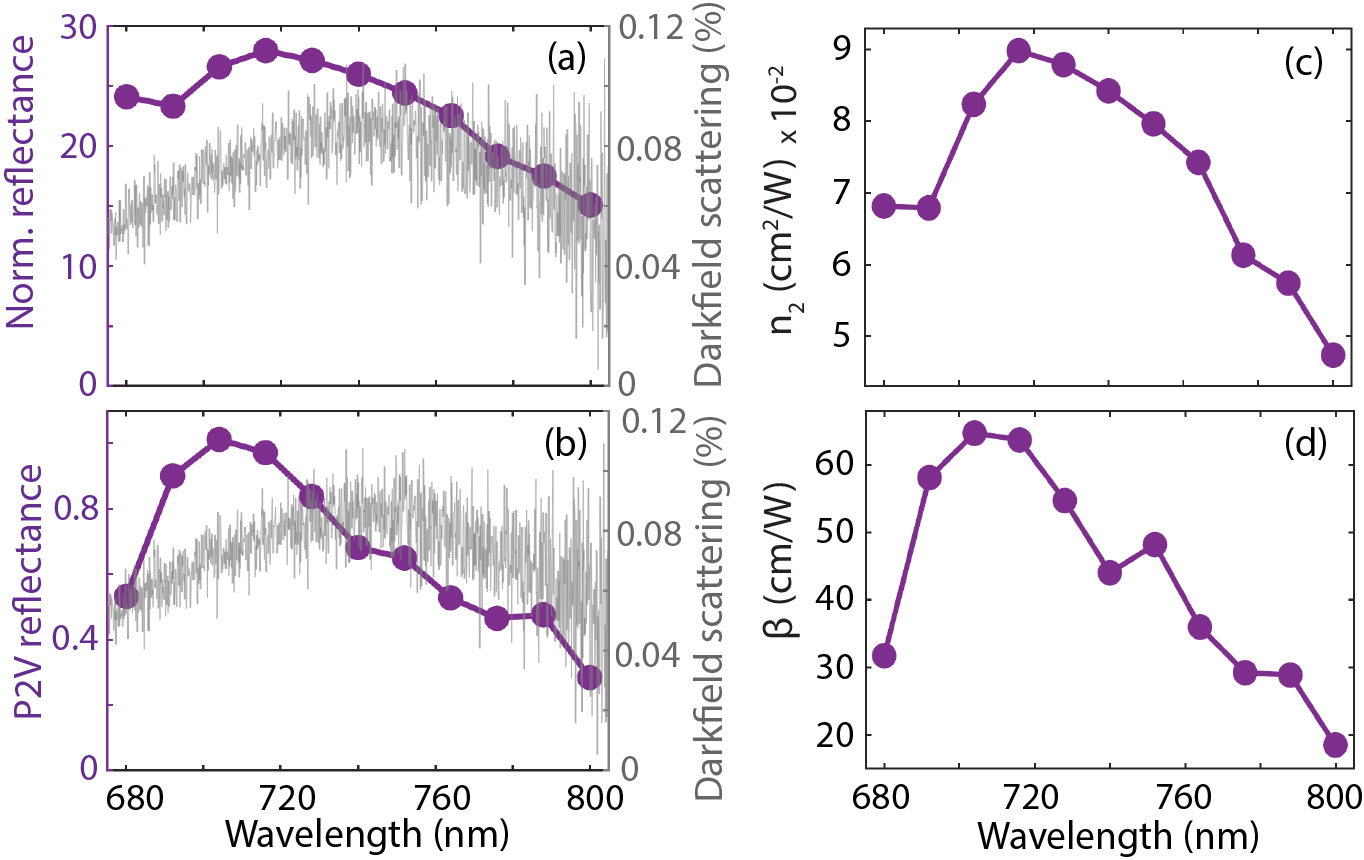}
\caption{Excitation wavelength scan for (a) OA and (b) CA at an average power of 900 pW. The grey solid curve is the extinction spectrum. (c, d) The extracted values of $n_2$ and $\beta$ at different wavelengths, respectively.}
\label{fig4}
\end{figure}

Furthermore, we investigate the effect of the cavity resonance on the nonlinear interaction using the methylene blue layer in the cavity. We scanned the pulsed laser wavelength from 680 to 800 nm at an average power of 900 pW. Indeed, the nonlinear response of the nanocavity is dependent on wavelength, as expected (Figure \ref{fig4}). In the OA measurements, the peak response occurs at 716 nm, while in the CA measurements, the peak is blue-shifted by approximately 20 nm. The darkfield extinction spectrum has a more redshifted peak at around 740 nm. We attribute the spectral difference to a phase shift of the confined field in the gap out-coupled to farfield\cite{lombardi2016anomalous}. The extinction spectrum probes the far-field radiation while the nonlinear interactions occur inside the cavity in the near field. Both open and closed-aperture experiments indicate that the nanocavity response is maximized at a specific wavelength and decreases as the wavelength moves away from this peak(Figures \ref{fig4} c,d). These strong dependences of the nonlinear parameters on wavelength show that the nanocavity resonance strongly determines the nonlinear interaction.


In conclusion, we use a few photons to extract the third-order nonlinear  parameters ($n_2$ and $\beta$) of nanomaterials inside plasmonic nanocavities. The few photons is sufficient to retrieve the parameters of different nanomaterials inside the nanocavity, namely Au and perovskite nano-objects and methylene blue monolayer. Particularly, we observe that the $n_2$ values are much higher for a nanocavity with a high field enhancement, indicating a strong modification of the reflection coefficient in the gap. Moreover, the $\beta$ indicates two-photon absorption and saturable absorption for the nanoparticles and the molecular layer, respectively.  Beyond the nanomaterials examined here, our approach is applicable to studying the nonlinear optical properties of delicate biomolecules and nanomaterials that can be easily damaged by high laser intensities. Moreover, a characterization of the nonlinear optical properties of nanomaterials can enable their applications in nonlinear optical devices such as optical switches, saturable absorbers, and optical limiters. Additionally, these nonlinear effects are also exciting for quantum experiments at few photon levels.

\begin{backmatter}
\bmsection{Acknowledgments} We thank Klaus Boller and Herman Offerhaus for useful feedback and stimulating discussions. We acknowledge Damien Hudry at the Institute for Microstructure Technology at KIT for leading the synthesis efforts on the perovskite nanocrystals and Melissa Goodwin from the MESA+ institute at UT for performing STEM measurements. 

\bmsection{Disclosures} The authors declare no conflicts of interest.

\bmsection{Funding} This publication is part of the project "Three seeds for the quantum/nano‐revolution" with project number NWA.1418.22.001 and the project OCENW.XS22.1.107, which are financed by the Dutch Research Council (NWO). Additionally, the Faculty of Science and Technology, University of Twente supported the project with a tenure-track start-up fund. We also gratefully acknowledge the financial support provided by NWO through the Knowledge and Innovation Covenant (KIC) program. This publication is part of the project Diffuse Irradiance Redirector for Efficient ConcenTration (DIRECT) (with project number KICH1.ED02.20.006 of the research programme NWO Kennis- en innovatieconvenant Innovations for Wind and Solar which is (partly) financed by the Dutch Research Council (NWO).

\bmsection{Data availability} Data underlying the results presented in this paper are available in Ref. [xx].

\bmsection{Supplemental document}
See Supplement 1 for supporting content. 

\end{backmatter}


\bibliography{Optica-journal-template}

\begin{thebibliography}{10}
\newcommand{\enquote}[1]{``#1''}

\bibitem{dini2016nonlinear}
D.~Dini, M.~J. Calvete, and M.~Hanack, {\protect\JournalTitle{Chemical reviews}} \textbf{116}, 13043 (2016).

\bibitem{wang2021highly}
F.~Wang, H.~Chen, D.~Lan, \emph{et~al.}, {\protect\JournalTitle{Advanced Optical Materials}} \textbf{9}, 2100795 (2021).

\bibitem{wang2019saturable}
G.~Wang, A.~A. Baker-Murray, and W.~J. Blau, {\protect\JournalTitle{Laser \& Photonics Reviews}} \textbf{13}, 1800282 (2019).

\bibitem{min2008all}
C.~Min, P.~Wang, C.~Chen, \emph{et~al.}, {\protect\JournalTitle{Optics letters}} \textbf{33}, 869 (2008).

\bibitem{sirleto2023introduction}
L.~Sirleto and G.~C. Righini, {\protect\JournalTitle{Micromachines}} \textbf{14}, 614 (2023).

\bibitem{koenderink2015nanophotonics}
A.~F. Koenderink, A.~Al{\`u}, and A.~Polman, {\protect\JournalTitle{Science}} \textbf{348}, 516 (2015).

\bibitem{xiang2016nanoparticle}
D.~Xiang and R.~Gordon, {\protect\JournalTitle{ACS Photonics}} \textbf{3}, 1421 (2016).

\bibitem{Wang_AdvOptPhoton_2011}
Y.~Wang, C.-Y. Lin, A.~Nikolaenko, \emph{et~al.}, {\protect\JournalTitle{Adv. Opt. Photon.}} \textbf{3}, 1 (2011).

\bibitem{singhal2010study}
H.~Singhal, R.~Ganeev, P.~Naik, \emph{et~al.}, {\protect\JournalTitle{Journal of Physics B: Atomic, Molecular and Optical Physics}} \textbf{43}, 025603 (2010).

\bibitem{bonacina2020harmonic}
L.~Bonacina, P.-F. Brevet, M.~Finazzi, and M.~Celebrano, {\protect\JournalTitle{Journal of Applied Physics}} \textbf{127} (2020).

\bibitem{shen2016two}
Y.~Shen, A.~J. Shuhendler, D.~Ye, \emph{et~al.}, {\protect\JournalTitle{Chemical Society Reviews}} \textbf{45}, 6725 (2016).

\bibitem{lu2022two}
X.~Lu, D.~Punj, and M.~Orrit, {\protect\JournalTitle{Nano Letters}} \textbf{22}, 4215 (2022).

\bibitem{tomita2006magneto}
S.~Tomita, T.~Kato, S.~Tsunashima, \emph{et~al.}, {\protect\JournalTitle{Physical review letters}} \textbf{96}, 167402 (2006).

\bibitem{huang2023weak}
J.-H. Huang, J.~S. Lundeen, K.~M. Jordan, \emph{et~al.}, {\protect\JournalTitle{Physical Review A}} \textbf{108}, 033724 (2023).

\bibitem{obermeier2018nonlinear}
J.~Obermeier, T.~Schumacher, and M.~Lippitz, {\protect\JournalTitle{Advances in Physics: X}} \textbf{3}, 1454341 (2018).

\bibitem{schumacher2011nanoantenna}
T.~Schumacher, K.~Kratzer, D.~Molnar, \emph{et~al.}, {\protect\JournalTitle{Nature communications}} \textbf{2}, 333 (2011).

\bibitem{driben2009low}
R.~Driben, A.~Husakou, and J.~Herrmann, {\protect\JournalTitle{Optics express}} \textbf{17}, 17989 (2009).

\bibitem{besner2006fragmentation}
S.~Besner, A.~V. Kabashin, and M.~Meunier, {\protect\JournalTitle{Applied Physics Letters}} \textbf{89} (2006).

\bibitem{samoc1998femtosecond}
M.~Samoc, A.~Samoc, B.~Luther-Davies, \emph{et~al.}, {\protect\JournalTitle{JOSA B}} \textbf{15}, 817 (1998).

\bibitem{sheik1990sensitive}
M.~Sheik-Bahae, A.~A. Said, T.-H. Wei, \emph{et~al.}, {\protect\JournalTitle{IEEE journal of quantum electronics}} \textbf{26}, 760 (1990).

\bibitem{van1997z}
E.~W. Van~Stryland and M.~Sheik-Bahae, \enquote{Z-scan technique for nonlinear materials characterization,} in \emph{Materials characterization and optical probe techniques: a critical review,} , vol. 10291 (SPIE, 1997), pp. 488--511.

\bibitem{tian2024dispersion}
X.~Tian, H.-s. Lu, T.~Qian, \emph{et~al.}, {\protect\JournalTitle{Applied Physics Letters}} \textbf{124} (2024).

\bibitem{maier2007plasmonics}
S.~A. Maier \emph{et~al.}, \emph{Plasmonics: fundamentals and applications}, vol.~1 (Springer, 2007).

\bibitem{kauranen2012nonlinear}
M.~Kauranen and A.~V. Zayats, {\protect\JournalTitle{Nature photonics}} \textbf{6}, 737 (2012).

\bibitem{butet2015optical}
J.~Butet, P.-F. Brevet, and O.~J. Martin, {\protect\JournalTitle{ACS nano}} \textbf{9}, 10545 (2015).

\bibitem{Bouhelier_PhysRevLett_2003}
A.~Bouhelier, M.~Beversluis, A.~Hartschuh, and L.~Novotny, {\protect\JournalTitle{Phys. Rev. Lett.}} \textbf{90}, 013903 (2003).

\bibitem{Lippitz_NanoLett_2005}
M.~Lippitz, M.~A. van Dijk, and M.~Orrit, {\protect\JournalTitle{Nano Letters}} \textbf{5}, 799 (2005). PMID: 15826131.

\bibitem{li_nonlinear_2017}
G.~Li, S.~Zhang, and T.~Zentgraf, {\protect\JournalTitle{Nature Reviews Materials}} \textbf{2} (2017).

\bibitem{ojambati2020efficient}
O.~S. Ojambati, R.~Chikkaraddy, W.~M. Deacon, \emph{et~al.}, {\protect\JournalTitle{Nano Letters}} \textbf{20}, 4653 (2020).

\bibitem{wang_NatRevChem_2022}
M.~Wang, T.~Wang, O.~S. Ojambati, \emph{et~al.}, {\protect\JournalTitle{Nature Reviews Chemistry}} pp. 1--24 (2022).

\bibitem{baumberg_extreme_2019}
J.~J. Baumberg, J.~Aizpurua, M.~H. Mikkelsen, and D.~R. Smith, {\protect\JournalTitle{Nature Materials}} \textbf{18}, 668 (2019).

\bibitem{liu2001theoretical}
X.~Liu, S.~Guo, H.~Wang, and L.~Hou, {\protect\JournalTitle{Optics communications}} \textbf{197}, 431 (2001).

\bibitem{ganeev2005reflection}
R.~Ganeev and A.~Ryasnyansky, {\protect\JournalTitle{physica status solidi (a)}} \textbf{202}, 120 (2005).

\bibitem{petrov1996reflection}
D.~Petrov, {\protect\JournalTitle{JOSA B}} \textbf{13}, 1491 (1996).

\bibitem{ganeev2008single}
R.~Ganeev, {\protect\JournalTitle{Applied Physics B}} \textbf{91}, 273 (2008).

\bibitem{Smith_JAppPhys_1999}
D.~D. Smith, Y.~Yoon, R.~W. Boyd, \emph{et~al.}, {\protect\JournalTitle{Journal of Applied Physics}} \textbf{86}, 6200 (1999).

\bibitem{ojambati2024few}
O.~S. Ojambati, K.~B. Arnard{\'o}ttir, B.~W. Lovett, \emph{et~al.}, {\protect\JournalTitle{Nanophotonics}} \textbf{13}, 2679 (2024).

\bibitem{lombardi2016anomalous}
A.~Lombardi, A.~Demetriadou, L.~Weller, \emph{et~al.}, {\protect\JournalTitle{ACS photonics}} \textbf{3}, 471 (2016).

\end{thebibliography}



\ifthenelse{\equal{\journalref}{aop}}{%
\section*{Author Biographies}
\begingroup
\setlength\intextsep{0pt}
\begin{minipage}[t][6.3cm][t]{1.0\textwidth} 
  \begin{wrapfigure}{L}{0.25\textwidth}
    \includegraphics[width=0.25\textwidth]{john_smith.eps}
  \end{wrapfigure}
  \noindent
  {\bfseries John Smith} received his BSc (Mathematics) in 2000 from The University of Maryland. His research interests include lasers and optics.
\end{minipage}
\begin{minipage}{1.0\textwidth}
  \begin{wrapfigure}{L}{0.25\textwidth}
    \includegraphics[width=0.25\textwidth]{alice_smith.eps}
  \end{wrapfigure}
  \noindent
  {\bfseries Alice Smith} also received her BSc (Mathematics) in 2000 from The University of Maryland. Her research interests also include lasers and optics.
\end{minipage}
\endgroup
}{}

\end{document}